\documentclass[seceq,preprint]{ptptex}

\preprintnumber{YITP-07-73}

\title{From inflation to dark energy in the non-minimal modified gravity}

\author{Shin'ichi \textsc{Nojiri}$^{*1}$, Sergei D. \textsc{Odintsov}\footnote{also at Lab. Fundam. Study, Tomsk State
Pedagogical University, Tomsk}$^{*2}$,
and Petr V. \textsc{Tretyakov}$^{*3}$}
\inst{$^{*1}$ Department of Physics, Nagoya University, Nagoya 464-8602. Japan \\
$^{*2}$ Instituci\`{o} Catalana de Recerca i Estudis Avan\c{c}ats (ICREA)
and Institut de Ciencies de l'Espai (IEEC-CSIC),
Campus UAB, Facultat de Ciencies, Torre C5-Par-2a pl, E-08193 Bellaterra
(Barcelona), Spain \\
$^{*3}$ Sternberg Astronomical Institute, Moscow 119992, Russia
}

\abst{
We consider the modified gravity non-minimally coupled with matter
Lagrangian for the description of early-time and late-time universe.
Such $F(R)$ ($F(G)$) gravity in the absence of non-minimal coupling is
viable theory which passes the local tests and reproduces the $\Lambda$CDM
era. For qualitatively similar choice of non-minimal gravitational
coupling function it is shown that the unified description of early-time
inflation and late-time cosmic acceleration is possible. It is interesting
that matter (scalar) which supports the inflationary era is
gravitationally screened at late times. Hence, it may be effectively
invisible at current universe.  }

\newcommand{\be}{\begin{equation}}
\newcommand{\ee}{\end{equation}}
\newcommand{\bea}{\begin{eqnarray}}
\newcommand{\eea}{\end{eqnarray}}
\newcommand{\beaa}{\begin{eqnarray*}}
\newcommand{\eeaa}{\end{eqnarray*}}

\newcommand{\nn}{\nonumber \\}
\newcommand{\e}{{\rm e}}

\begin{document}

\maketitle

\section{Introduction}

It has been realized recently that modified gravity (for a review, see
\citen{review}) may describe the universe expansion history quite
realistically. In such a picture, the Einstein gravity is just the
approximation to more complete (classical) gravity where some
gravitational terms which are leading at early-time universe support the
inflation while other terms which are dominant at late-time universe cause
the cosmic acceleration. The simple gravitational model of such
unification of
inflation with late-time acceleration is suggested in
 ref.\citen{NOPRD}. It also has been discovered the viable modified
 gravity\cite{hu} (for related models, see \citen{nohs,UU,other}) which
describes the $\Lambda$CDM epoch similar to usual Einstein gravity with
cosmological constant. Moreover, such theory passes the local
tests\cite{hu,nohs} (for review of confronting the observational data with
modified gravity predictions, see \citen{salvo}). Finally, such class of
modified gravities may successfully describe the universe expansion
history from the early-time inflation till late-time
acceleration\cite{nohs,UU} with correct intermediate epoch.

Another gravitational source of the inflation and dark energy
may be the non-minimal coupling of some geometrical invariants
function with matter Lagrangian. Such non-minimal modified gravity
has been introduced in refs. \citen{noplb,allemandi} for the study
of gravity assisted dark energy occurrence. It may be also applied
for realization of dynamical cancellation of cosmological
constant\cite{constant}. The viability criteria for such theory
was recently discussed in refs. \citen{lobo, faraoni, bertolami}. In
the present work we consider non-minimal modified gravity for the
class of functions introduced in ref. \citen{hu,UU}. For simplicity, as
the matter we consider just usual scalar kinetic term. It is shown
that for such model one can easily achieve the unification of the
inflation with cosmic acceleration. Moreover, it may suggest the
explanation why inflaton is not seen at the late universe: because
it is screened by gravitational function which quickly goes to
zero with universe expansion.

 \section{Unification of the inflation with late-time
acceleration in the non-minimal modified gravity}

In this section we obtain the equations of motion for general
modified gravity non-minimally coupled with matter. These
equations will be used for the investigation of the unified
inflation-late-time acceleration epoch emergence for several
realistic models.

The starting theory is:
\begin{equation}
S=\int d^4 x \sqrt{- g} \left [ \frac{1}{\kappa^2} R - f_1(A) L_d -f_2(A)\right ]\ ,
\label{1.1}
\end{equation}
where $A$  is some function of geometrical invariants and $L_d$ is
the matter Lagrangian. In this work we choose $A$ to be equal to
$R$ or Gauss-Bonnet invariant  $G$, for simplicity. We also assume
$L_d$ is scalar function: $L_d=\frac{1}{2}n^{-2}\dot \phi ^2,$
where $\phi = \phi (t)$ and metric corresponds to spatially-flat
FRW universe:
\begin{equation}
g_{\mu\nu}={\rm diag}(-n(t)^2, a(t)^2 , a(t)^2 , a(t)^2)\ .
 \label{1.2}
\end{equation}
Varying (\ref{1.1}) on $\phi$ one arrives to the scalar field
equation\cite{noplb}: $f_1(A)a^3\dot \phi n^{-1}=q={\rm const.}$, or
\begin{equation}
\dot \phi=f_1(A)^{-1}a^{-3} nq\ .
\label{1.4}
\end{equation}
Substituting it to (\ref{1.1}) one gets
\begin{equation}
S=\int d^4 x  \left [ \frac{1}{\kappa^2}Rna^3 - \frac{n
q^2}{2f_1(A)a^3} -f_2(A)na^3 \right ]=S_R+S_1+S_2\ .
\label{1.6}
\end{equation}
Varying (\ref{1.6}) on $n$ the FRW equation follows:
\be
\label{A1}
\delta S_R = \frac{6}{\kappa^2} \int d^4 x a\dot a^2  \delta n\ .
\ee
(after variation we may put $n=1$)
\bea
\label{A2}
\delta S_1 &=& \int d^4 x  \left [ -\frac{ q^2}{2f_1(A)a^3}\delta n + \frac{
 q^2}{2a^3}\frac{1}{f_1(A)^2}\frac{\partial f_1}{\partial A}\frac{\partial A}{\partial n}\delta n + \frac{
 q^2}{2a^3}\frac{1}{f_1(A)^2}\frac{\partial f_1}{\partial A}\frac{\partial A}{\partial \dot n}\delta \dot n
\right ] \nn
&=& \int d^4 x  \left [
  -\frac{ q^2}{2f_1a^3} + \frac{
 q^2}{2a^3}\frac{f_1'}{f_1^2} A_n + \frac{3}{2}\frac{
 q^2}{a^3}H\frac{f_1'}{f_1^2} A_{\dot n} + \frac{
 q^2}{a^3}\frac{(f_1')^2}{f_1^3} A_{\dot n}\dot A  \right. \nn
&& \left. -\frac{ q^2}{2a^3}\frac{f_1''}{f_1^2}A_{\dot n}\dot A -\frac{
 q^2}{2a^3}\frac{f_1'}{f_1^2}\frac{d A_{\dot n}}{d t}
\right ]\delta n \ ,\nn
\delta S_2 &=& -\int d^4x \left [a^3 f_2 +f_2'a^3A_n -
3a^3Hf_2'A_{\dot n} - a^3f_2'\frac{d A_{\dot n}}{dt}
-a^3f_2''A_{\dot n}\dot A \right ]\delta n \ .
\eea
Now FRW equation may be written in the following form:
\begin{equation}
\frac{6}{\kappa^2}H^2=\rho_{1}+\rho_{2}\ ,
\label{1.7}
\end{equation}
where
\bea
\label{1.8}
\rho_{1} &=& \frac{ q^2}{f_1a^6}    \left [
  \frac{1}{2} - \frac{1}{2}\frac{f_1'}{f_1}A_n - \frac{3}{2}H\frac{f_1'}{f_1} A_{\dot n} - \frac{(f_1')^2}{f_1^2} A_{\dot n}\dot A
  +\frac{1}{2}\frac{f_1''}{f_1}A_{\dot n}\dot A +\frac{1}{2}\frac{f_1'}{f_1}\frac{d A_{\dot n}}{d t} \right ]\ , \\
\label{1.9}
\rho_{2}&=& f_2 +f_2'A_n - 3Hf_2'A_{\dot n} - f_2'\frac{d A_{\dot n}}{dt} -f_2''A_{\dot n}\dot A\ .
\eea
The important remark is in order. Our FRW-like
equation contains higher derivative term which is multiplied by some
function. If this function approaches to zero during evolution we
discover so-called finite time singularity (for their classification, see \citen{tsujikawa}).
To avoid the singularity it is necessary that
this function (multiplying by higher derivatives)
is not equal to zero in any moment. It is clear from Eqs.
(\ref{1.8})-(\ref{1.9}) that higher derivatives appear only in
$\dot A$ term. Hence, the higher derivatives function is given by:
\begin{equation}
HD=\dot A A_{\dot n} \left [ \frac{
q^2}{f_1^2a^6}(\frac{1}{2}f_1'' - \frac{(f_1')^2}{f_1}) -f_2''
\right].
\label{1.10}
\end{equation}
When it is not zero, the finite-time singularity does not occur.

Let us now consider the following choice of function $A$: $A=R$.
Then $R=\frac{6}{n^2}\left[ \frac{\ddot a}{a} - \frac{\dot a}{a}\frac{\dot n}{n}+\frac{\dot a^2}{a^2} \right ]$,
$R_n=-2R=-12(\dot H+2H^2)$, $R_{\dot n}=-6H$.

The effective energy-density becomes (below $n=1$):
\bea
\label{2.1}
\rho_{1} &=& \frac{ q^2}{f_1a^6}    \left [
  \frac{1}{2} +3\frac{f_1'}{f_1}(\dot H + 7H^2) +6H \frac{(f_1')^2}{f_1^2} \dot R
  -3H\frac{f_1''}{f_1}\dot R  \right ]\ , \\
\label{2.2}
\rho_{2} &=& f_2 -6f_2'(\dot H +H^2) +6 H f_2''\dot R\ .
\eea
This FRW equation is used in the explicit analysis below.

Einstein gravity non-minimally coupled with matter. Let us
consider the Einstein gravity non-minimally coupled with matter
Lagrangian:
\be
\label{AA1}
S=\int d^4 x \sqrt{-\mathrm{g}} \left [
\frac{1}{\kappa^2}R - f_1(A) L_d\right ]\ ,
\ee
where it is chosen $f_2\equiv 0$. Note that recently observational bounds
for such non-minimal coupling were discussed in \citen{bertolami}.
Let us take the following explicit choice for function $f_1$ which
was discussed recently for modified $F(R)$ gravity in \citen{hu,nohs}
\begin{equation}
f_1=\frac{c_1R^k}{c_2R^k+1},
 \label{2.3}
\end{equation}
where $c_1$ and $c_2$ are the arbitrary constants.
The FRW equation looks as: $ \frac{6}{\kappa^2}H^2=\rho_1$.

Let us discuss the cosmologically important limit of this theory
$R\rightarrow 0$ which corresponds to current universe.
In this limit, the approximated FRW equation is obtained as follows:
\begin{equation}
\frac{6}{\kappa^2}\!H^2\!\!=\!\!\frac{18q^2}{a^6c_1 \!(6\dot
H\!\!+\!\!12H^2)^{\!k+2}}[\dot H^2(k+\!1)+H^4(k+\!14k)+\dot
HH^2(4+\!13k+\!4k^2)+k(k+\!1)H\ddot H].
\label{2.8}
\end{equation}
One can study the power-like  solutions of (\ref{2.8}):
$a=a_0t^x$, $H=\frac{x}{t}$, $\dot H=-\frac{x}{t^2}$, $\ddot H=2\frac{x}{t^3}$.
Substituting these
relations into (\ref{2.8}) we find: $ x=\frac{k+1}{3}$.
 From another side $a\propto t^{\frac{2}{3(w+1)}}$,
where $w$ is defined as $p=w\rho$, and therefore we find
$w=\frac{1-k}{1+k}$. Hence, late-time accelerated universe
($-1<w<-\frac{1}{3}$) occurs when $k>2$. Effective phantom regime
emerges when $k<-1$\cite{tsujikawa}.
When $w$ is negative, the universe shrinks when $t$ is positive since $x<0$.
If we replace $t$ with $t_s - t$ and assume $t<t_s$, the universe is expanding.
The universe has a singlarity at $t=t_s$, which corresponds to so-called
``Big Rip'' singularity.
We should also note that exact
de Sitter solution is impossible in this theory because there is scale
factor $a$ in equation (\ref{2.8}). It always depends on time
while all other parameters are constant. Nevertheless, de Sitter space may
be
realized asymptotically when $k=-1$.

At the early universe it is known that $R\rightarrow \infty$.  In
this case $f_1\simeq\frac{c_1}{c_2}-\frac{B}{R^k}$, where $B$ is
some constant. Then $\rho_1$  is about $\rho_1\sim f_1^{-1}a^{-6}$
where $f_1^{-1}\simeq \frac{c_2}{c_1} + R^{-k}/c_1$ at this
period. Hence, one  sees that in this theory inflation (quasi-de Sitter
stage) is possible in principal, because $\rho_1$ is sufficiently
large at early times. The inflation stage is not stable as it should
be. Indeed, $a$ increases exponentially while $f_1$
decreases only as some power of time. Thus, the principal
possibility to unify the inflation with late-time acceleration in
above theory is proved. The important property of such unification
is quick decay of gravitationally-coupled inflaton. That may
suggest the scenario why the early time inflaton is not seen in
current universe: it may be screened by gravitational coupling
function which tends to zero with curvature decrease.

Modified $F(R)$ gravity non-minimally coupled with matter. Let us
start from the theory (\ref{1.1}) where $f_1$ is given by
(\ref{2.3})  and $f_2$ is taken in the similar form:
\begin{equation}
f_2=\frac{c_3R^m}{c_4R^m+1}.
 \label{2.4}
\end{equation}
The unification of early-time inflation with late-time acceleration in
such theory without non-minimal coupling with matter has been recently
studied in ref. \citen{UU}. Let us estimate now the role of non-minimal
coupling with matter Lagrangian.

At late-time universe when $R\rightarrow 0$ the cosmic
acceleration was studied in ref. \citen{hu} in the theory without
non-minimal coupling ($f_1=0$). Moreover, it was demonstrated
there that the late-time universe corresponds to usual $\Lambda$CDM
epoch subject the corresponding choice of parameters of the
theory. The theory also passes local tests\cite{hu,nohs}. Hence,
when the parameters of the theory are chosen in such a way that
$\rho_1$ may be neglected if compare with $\rho_2$ then late-time
cosmic acceleration corresponds to $\Lambda$CDM epoch. The dark
energy universe is realistic and complies with observational data.
In the opposite situation, when $\rho_1$ gives the leading
contribution to the effective energy-density one comes back to the
case described above for pure Einstein gravity non-minimally
coupled with matter where late-time acceleration is again
possible.

At the early-time universe
 we have $f_2=\frac{c_3}{c_4}-\frac{B}{R^m}$, where $B$
is some constant and respectively $ \rho_2=C-\frac{B_1}{R^m}$,
where $C$ and $B_1$ are constants. When
 only $\rho_2$ presents in FRW equation, we have quasi-inflationary
 stage as the
solution. From another side, one may estimate
$\rho_1$  as $\rho_1\sim f_1^{-1}a^{-6}$ at this period.
 Since inflation starts
from some $a_*\neq 0$, it is clear that $\rho_1$ makes some contribution
to the inflation. However, this contribution very rapidly decreases
because
$a$ increases exponentially while $f_1$ decreases only as some
power of time. Thus, we again have the theory with unified inflation
and late-time acceleration. It is important to note here that the presence
of non-minimal coupling with matter (after some fine-tuning) may improve
the theoretical estimations which are compared with the observational data
like cosmological parameters (effective equation of state parameter, etc),
and  cosmological perturbations.

Another choice for functions $f_1$, $f_2$ is motivated by the
realistic and viable model which was considered in ref. \citen{UU}:
\bea
f_2 (R) &=&\frac{(R-R_0)^{2k+1}+R_0^{2k+1}}{c_3+c_4\left[
(R-R_0)^{2k+1} + R_0^{2k+1} \right]}\ ,
 \label{2.9} \\
f_1 (R) &=& \frac{(R-R_0)^{2k+2m+1}+R_0^{2k+2m+1}}{c_1+c_2\left[
(R-R_0)^{2k+2m+1}+R_0^{2k+2m+1} \right]}\ .
 \label{2.10}
\eea
It is known\cite{UU} that the theory with $f_1=0$  leads to the
unification of the early-time inflation  and late-time acceleration with
realistic universe expansion history(radiation/matter dominance,
transition from deceleration to acceleration) between these two eras.
This could be realized since $f_2(R)$ satisfies the following conditions:
\be
\label{Uf1}
\lim_{R\to\infty} f_2 (R) = \Lambda_i\equiv \frac{1}{c_4}\ ,
\quad f_2 (R_0)= 2R_0 = \frac{R_0^{2k+1}}{c_3 + c_3 R_0^{2k+1}} \ ,\quad f_2'(R_0)\sim 0\ .
\ee
Here $R_0$ is current curvature  $R_0\sim \left(10^{-33}{\rm eV}\right)^2$.
Then in the above model, the universe starts from the inflation
driven by the effective cosmological constant $\Lambda_i$ at the early stage,
where curvature is very large. As curvature becomes smaller, the
effective cosmological constant also becomes smaller. After that the
radiation/matter dominates.
When the density of the radiation and the matter becomes small and the
curvature goes to the value $R_0$, there  appears the small effective cosmological constant
$2R_0$. Hence, the current cosmic expansion could start.
We also note that $f_2(R)$ satisfies the condition
\be
\label{Uf4}
\lim_{R\to 0} f(R) = 0\ ,
\ee
which shows the existence of the flat spacetime solution.

Now let us check at which conditions  $f_1$ term (almost) does not change late-time
acceleration. To satisfy this, it is necessary
 that $\rho_1$ tends to zero at $R\rightarrow R_0$ more rapidly than
$\rho_2$ tends to constant.  Simple check shows that acceptable
choice of parameters corresponds to $c_1=-c_2R_0^{2k+2m+1}$. For
$m>0$ there is no any influence on late-time acceleration from
scalar term, because near $R=R_0$ one finds $\rho_2\sim const
+(R-R_0)^{2k+1}$, while $\rho_1\sim (R-R_0)^{2k+2m+1}$. From
another side, there is some contribution from scalar term to
early-time inflation and, therefore, to cosmological perturbations.
However, this
contribution very rapidly decreases as $\frac{1}{a^6}$.
 Moreover, if $2k+2m+1>1$
 there may be one more accelerating regime in the future.
Thus, it is shown the principal possibility to unify the inflation with
late-time acceleration in the modified gravity model\cite{UU} even in the
presence of non-minimal coupling with matter.

\section{Inflation and late-time acceleration in non-minimal
 F(G)-gravity}

Let us study the theory with $A=G$, where $G =
R^2-4R_{\mu\nu}^2+R_{\mu\nu\alpha\beta}^2$ -- is Gauss-Bonnet
invariant. Such F(G) gravity has been introduced in ref. \citen{G} as
gravitational alternative for dark energy (for recent study of
local tests in such theory with power-law function F(G),
see\cite{barrow}). The dark energy application of
its non-minimal coupling with
matter was investigated in ref. \citen{not}.

Let us consider the same qualitative choice for functions $f_1$,
$f_2$ as in previous section:
\begin{equation}
f_1(G)=\frac{G^k}{c_1G^k+c_2}\ , \quad
f_2(G)=\frac{G^m}{c_3G^m+c_4}\ .
\end{equation}
It is known\cite{UU} that in the absence of non-minimal coupling
$f_1=0$ such a model naturally leads to unification of the
inflation with late-time acceleration being viable theory. The
presence of $f_1$ term does not qualitatively change the situation
at the early universe. Indeed, when $t\rightarrow 0$,
$\rho_{G2}\simeq const$ while $\rho_{G1}\propto a^{-6}$. Hence,
$\rho_{G1}$ rapidly decreases at the expanding universe and can be
neglected. The inflationary  epoch emerges. At the
late-time universe ($t\rightarrow \infty$) one can see that
$\rho_{G2}\rightarrow 0$ supporting the acceleration, while
$\rho_{G1}\propto a^{-6}G^{-k}$. Hence, for sufficiently big $k$,
we may get the situation where $\rho_{G1}\rightarrow const$ or it
is growing. In this case, the dominant contribution to the
late-time acceleration is due to non-minimal coupling with matter.
In this case, the matter perturbations of such theory may be significally
changed if compare with the case without non-minimal coupling.

Let us discuss the occurrence of cosmic acceleration in more
detail. In this case one can put $\rho_{G2}=0$ and $f_1=G^k/c_2$.
Searching for the power-like solutions:
$a=a_0t^x$, $H=\frac{x}{t}$, $\dot H=-\frac{x}{t^2}$, $\ddot H=2\frac{x}{t^3}$,
$G\propto t^{-4}$, $f_1\propto t^{-4k}$ in FRW Eq.(7), we find:
$-2=-6x+4k$ or $x=\frac{2k+1}{3}$, Thus, the effective
equation of state parameter is given by $w=\frac{1-2k}{1+2k}$. The
late-time accelerating universe ($-1<w<-\frac{1}{3}$) occurs if
$k>1$. Effective phantom era emerges
if $k<-\frac{1}{2}$ while at the very large $k$ the universe is described
by
$\Lambda$CDM cosmology. The case of $k=-\frac{1}{2}$ leads to
asymptotically de Sitter space. Thus, the principal possibility of the
unification of the early-time inflation with late-time
acceleration is again established. The realistic intermediate epoch
follows in the same way as it is shown in ref. \citen{UU}.

\section{ Discussion}

In summary, we considered $f(R)$ or $F(G)$ modified gravity non-
minimally coupled with matter Lagrangian which is chosen to be the scalar
kinetic energy, for simplicity. It is demonstrated that in such theory
with the special choice of gravitational functions the realistic universe
expansion history emerges quite naturally. In particulary, the unification
of the early-time inflation with late-time acceleration occurs within the
same theory. In addition, as non-minimal gravitational coupling with
matter goes to zero at  late times, there appears the scenario
explaining why scalars may be not seen in the present universe. Moreover, one can
conjecture that it is really modified gravity which is responsible for
early-time inflation. The success of some scalar models of inflation with
specific potentials may be explained by occasional reason, i.e. by the fact
that non-minimal gravitational coupling at very early universe is
approximately constant.

It is not hard to extend our formulation for more complicated theories.
For instance, one can include the scalar potential into consideration in
matter Lagrangian. Other fields like spinors or vectors may be considered in
the similar fashion of non-minimal gravitational coupling
with matter Lagrangian. From another point, other types of realistic
modified gravity may be investigated: the model unifying $R^m$ inflation
with $Lambda$CDM epoch \cite{R} or non-local gravity
\cite{woodard,nonlocal}. The non-minimal versions of such models will be
investigated elsewhere.

\section*{Acknowledgements.}

The research by S.N. has been supported in part by the
Monbusho
grant no.18549001
and 21st Century COE Program of Nagoya University
provided by JSPS (15COEG01).
The research by S.D.O. has been supported in part by the projects
FIS2006-02842, FIS2005-01181 (MEC, Spain), by RFBR grant 06-01-00609
(Russia) and by YITP, Kyoto.

\appendix

\section{Classically equivalent forms of non-minimal modified gravity}

One can rewrite the action (\ref{1.1}) with $A=R$ by using
 the auxilliary field(s).
First we introduce two scalar field $\zeta$ and $\eta$ and
rewrite (\ref{1.1}) as \cite{noplb}
\be
\label{LR18}
S=\int d^4 x \sqrt{-g}\left\{{1 \over \kappa^2}\zeta - f_1(\zeta) L_d  - f_2(\zeta)
+ \eta \left(R - \zeta\right) \right\}\ .
\ee
Using the equation $\zeta=R$ given by the variation over $\eta$,
the action  (\ref{LR18}) is reduced into the original one  (\ref{1.1}).
Varying over $\zeta$, we obtain
\be
\label{LR19}
\eta = {1 \over \kappa^2} - f_1'(\zeta) L_d - f_2'(\zeta)\ .
\ee
By substituting (\ref{LR19}) and deleting $\eta$ in (\ref{LR18}), one gets
\be
\label{LR20}
S=\int d^4 x \sqrt{-g}\left\{\left( {1 \over \kappa^2} - f_2'(\zeta) \right) R
 - \left( f_1'(\zeta) R + f_1(\zeta) \right) L_d
 - f_2(\zeta) - f_2'(\zeta) \zeta + {1 \over \kappa^2}\zeta \right\}\ .
\ee
By rescaling the metric as $g_{\mu\nu}\to \e^\sigma g_{\mu\nu}$ with
\be
\label{BBB1}
\e^{-\sigma} = 1 - \kappa^2 f_2'(\zeta) \ ,
\ee
 the Einstein frame action follows\cite{faraoni,noplb}:
\bea
\label{BBB2}
S&=&\int d^4 x \sqrt{-g}\left[\frac{1}{\kappa^2}\left(R
 - \frac{3}{2}g^{\mu\nu}\partial_\mu\sigma \partial_\nu\sigma\right) \right. \nn
&& - \left\{ f_1'(\zeta(\sigma)) \e^{\sigma} \left( R - 3 \Box \sigma
 - \frac{3}{2} \partial_\mu \sigma \partial^\mu \sigma \right)  + \e^{2\sigma} f_1(\zeta(\sigma)\right\}
L_d\left(\e^\sigma g_{\mu\nu}, \phi\right) \nn
&& \left.  - \e^{2\sigma} f_2(\zeta(\sigma)) + \e^\sigma \zeta(\sigma)\right]\ .
\eea
Here we have solved (\ref{BBB1}) with respect to $\zeta$ as $\zeta=\zeta(\sigma)$.
Such the non-linear action includes Brans-Dicke type scalar $\sigma$
and the scalar $\phi$
corresponding to the dark energy, that is,  two scalars appear.
Such classical equivalence of two formulations does not mean the physical
equivalence as is explained in \cite{troisi,faraoni1}.


\begin{thebibliography}{99}

\bibitem{review}
S.~Nojiri and S.~D.~Odintsov,
arXiv:hep-th/0601213;
J.\ Phys.\ Conf.\ Ser. {\bf 66}, 012005 (2007)
[arXiv:hep-th/0611071].

\bibitem{NOPRD}
S.~Nojiri and S.~D.~Odintsov,
Phys.\ Rev.\ D {\bf 68}, 123512 (2003)
[arXiv:hep-th/0307288].

\bibitem{hu}
W.~Hu and I.~Sawicki,
arXiv:0705.1158[astro-ph].

\bibitem{nohs}
S.~Nojiri and S.~D.~Odintsov,
Phys.\ Lett.\ B {\bf 652}, 343 (2007),
[arXiv:0706.1378].

\bibitem{UU}
S.~Nojiri and S.~D.~Odintsov,
arXiv:0707.1941[hep-th].

\bibitem{other}
S.~Appleby and R.~A.~Battye,
arXiv:0705.3199[astro-ph];
L.~Pogosian and A.~Silvestri,
arXiv:0709.0296[astro-ph];
S.~Tsujikawa,
arXiv:0709.1391[astro-ph].

\bibitem{salvo}
S.~Capozziello and M.~Francaviglia,
arXiv:0706.1146[astro-ph].

\bibitem{noplb}
S.~Nojiri and S.~D.~Odintsov,
Phys.\ Lett.\ B {\bf 599}, 137 (2004)
[astro-ph/0403622];
PoS {\bf WC2004}, 024 (2004)
[arXiv:hep-th/0412030].

\bibitem{allemandi}
G.~Allemandi, A.~Borowiec, M.~Francaviglia, and S.~D.~Odintsov,
Phys.\ Rev.\ D {\bf 72}, 063505 (2005)
[arXiv:gr-qc/0504057].

\bibitem{constant}
S.~Mukohyama and L.~Randall,
Phys.\ Rev.\ Lett. {\bf 92}, 211302 (2004);
T.~Inagaki, S.~Nojiri and S.~D.~Odintsov,
JCAP {\bf 0506}, 010 (2005)
[arXiv:gr-qc/0504054];
A.~D.~Dolgov and M.~Kawasaki,
arXiv:astro-ph/0307442.

\bibitem{lobo}
O.~Bertolami, C.~Boehmer, T.~Harko, and F.~Lobo,
Phys.\ Rev.\ D {\bf 75}, 104016 (2007)
[arXiv:0704.1733];
T.~Koivisto,
Class.\ Quant.\ Grav.\ {\bf 23}, 4289 (2006)
[arXiv:gr-qc/0505128].

\bibitem{faraoni}
V.~Faraoni,
arXiv:0710.1291[gr-qc].

\bibitem{bertolami}
O.~Bertolami and J.~Paramos,
arXiv:0709.3988[astro-ph].

\bibitem{tsujikawa}
S.~Nojiri, S.~D.~Odintsov, and S.~Tsujikawa,
Phys.\ Rev.\ D {\bf 71}, 063004 (2005)
[arXiv:hep-th/0501025].

\bibitem{G}
S.~Nojiri and S.~D.~Odintsov,
Phys.\ Lett.\ B {\bf 631}, 1 (2005)
[arXiv:hep-th/0508049].

\bibitem{barrow}
G.~Cognola, E.~Elizalde, S.~Nojiri, S.~D.~Odintsov, and S.~Zerbini,
Phys.\ Rev.\ D {\bf 73}, 084007 (2006)
[arXiv:hep-th/0601008];
B.~Li, J.~Barrow and D.~Mota,
Phys.\ Rev.\ D {\bf 76}, 044027 (2007)
[arXiv:0705.3795];
S.~Davis,
arXiv:0709.4453[hep-th].

\bibitem{not}
S.~Nojiri, S.~D.~Odintsov, and P.~Tretyakov,
Phys.\ Lett.\ B {\bf 651}, 224 (2007)
[arXiv:0704.2520].

\bibitem{R}
S.~Nojiri and S.~D.~Odintsov,
arXiv:0710.1738[hep-th].

\bibitem{woodard}
S.~Deser and R.~Woodard,
Phys.\ Rev.\ Lett.\ {\bf 99}, 111301 (2007).

\bibitem{nonlocal}
S.~Nojiri and S.~D.~Odintsov,
arXiv:0708.0924[hep-th].

\bibitem{troisi}
S.~Capozziello, S.~Nojiri, S.~D.~Odintsov, and A.~Troisi,
Phys.\ Lett.\ B {\bf 639}, 135 (2006) 
[arXiv:astro-ph/0604431];
S.~Nojiri and S.~D.~Odintsov, 
Phys.\ Rev.\ D {\bf 74}, 086005 (2006)
[arXiv:hep-th/0608008].

\bibitem{faraoni1}
V.~Faraoni and S.~Nadeau, 
Phys.\ Rev.\ D {\bf 75}, 023501 (2007).

\end{thebibliography}
\end{document}